\documentclass[aps,pra,twocolumn,10pt]{revtex4-1}

\usepackage{natbib}
\usepackage{amsmath,amssymb,bm}
\usepackage{dsfont}
\usepackage{graphics}

\usepackage{color}

\newcommand{\tr}{\text{tr}}
\newcommand{\ket}[1]{\left|{#1}\right\rangle}
\newcommand{\bra}[1]{\left\langle{#1}\right|}
\newcommand{\braket}[2]{\langle{#1}|{#2}\rangle}
\newcommand{\braketa}[2]{\left\langle \left.{#1}\right|{#2}\right\rangle}
\newcommand{\braketb}[2]{\left\langle {#1}\left|{#2}\right.\right\rangle}
\newcommand{\Dcirc}{{\cal D}^{\circ}}
\newcommand{\dbar}{d\hspace*{-0.08em}\bar{}\hspace*{0.1em}}

\begin{document}

\title{Parametric down-conversion beyond the semi-classical approximation}
\author{Filippus S. \surname{Roux}}
\email{froux@nmisa.org}
\affiliation{National Metrology Institute of South Africa, Meiring Naud{\'e} Road, Brummeria 0040, Pretoria, South Africa}

\begin{abstract}
Using a perturbative approach, we investigate the parametric down-conversion process without the semi-classical approximation. A Wigner functional formalism, which incorporates both the spatiotemproal degrees of freedom and the particle-number degrees of freedom is used to perform the analysis. First, we derive an evolution equation for the down-conversion process in the nonlinear medium. Then we use the perturbative approach to solve the equation. The leading order contribution is equivalent to the semi-classical solution. The next-to-leading order contribution provides a solution that includes the evolution of the pump field as a quantum field.
\end{abstract}

\maketitle

\section{\label{intro}Introduction}

Parametric down-conversion (PDC) is one of the most versatile processes used for preparing quantum states in quantum optics \cite{klyshko,hongmandel,shihrev,pdcspatcor}. The conservation of momentum and energy in the nonlinear process leads to entanglement in the spatiotemporal degrees of freedom \cite{arnaut,mair,torres1}, which is used in many quantum information applications, ranging from quantum key distribution \cite{gibson,zeil2006,walborn,vallone,qkdturb}, quantum teleportation \cite{teleoam,telezeil}, quantum ghost imaging \cite{ghost1,ghostph}, and quantum synchronization protocols \cite{giovan,homsync2}.

The PDC process also produces entanglement in the particle-number degrees of freedom, which becomes manifest in the production of squeezed states \cite{sqlight,leuchs1,bsqvac,squreview}. Applications of squeezed states include continuous variable teleportation \cite{cvteleport,advtelport}, quantum imaging \cite{gatti,brambilla,lloydqill}, quantum state engineering \cite{quaneng}, and quantum metrology \cite{giovannetti}.

In view of these two different aspects of the PDC process, a comprehensive analysis of the process in terms of both spatiotemporal degrees of freedom and particle-number degrees of freedom would provide a clearer picture of the process and the state it produces \cite{bright}. Analyses of this nature often impose a variety of assumptions and approximations to make them tractable \cite{sharapova,wasil,leuchs1,bsqvac}. One approximation that features prominently is the semi-classical approximation where the pump field is assumed to be an undepleted classical field. However, the preparation of high fidelity squeezed states requires intensive pumping of the nonlinear medium to increase the efficiency of the down-conversion process. In such cases, the validity of the semi-classical approximation becomes questionable \cite{deplete}.

It is challenging to perform the analysis without the semi-classical approximation, because in such a case, the Hamiltonian interaction term for PDC contains a product of three ladder operators. Commutations of operators consisting of a product of more than two ladder operators increase the number of ladder operators per term in the result. Consequently, such an analysis would involve an endless cascade of products of ladder operators.

In quantum field theory, this issue is mitigated with the aid of perturbation theory \cite{peskin}. It is assumed that the coupling constant associated with the interaction term is small. Therefore, in theoretical particle physics, one can expand the calculation in progressively more complicated Feynman diagrams representing higher order contributions to the process under investigation. Unfortunately, in the current context of highly efficient PDC, one cannot assume that the coupling constant is small enough to allow such a perturbative expansion.

Here, we investigate the PDC process under general conditions with the aid of a perturbative approach but using a different expansion parameter. Assuming that the pump is represented by an intense coherent state, one can argue that the representation of the coherent state on phase space gives a minimum uncertainty width that is much smaller than its distance from the origin. We use the relative width of the pump distribution as the expansion parameter in our perturbative analysis. Such a perturbative expansion becomes more accurate as the intensity of the pump increases.

To perform an analysis where both spatiotemporal degrees of freedom and particle-number degrees of freedom are addressed, one requires a more powerful formalism than current methods allow. The incorporation of all these degrees of freedom naturally leads to a functional formalism. Such a functional formalism has been developed recently \cite{mrowc,stquad,stquaderr,recent} and is currently being applied to investigate complex phenomena in quantum optics \cite{ipfe,entpdc}.

Here, we use the Wigner functional formalism \cite{stquad,stquaderr,wigfunk} to derive an evolution equation for the spontaneous parametric down-converted state that is produced by pumping a second-order nonlinear medium. We follow the same infinitesimal propagation approach that was introduced for single-photon states \cite{ipe,lindb,notrunc} and was later extended to multi-photon states in the context of atmospheric scintillation due to turbulence \cite{ipfe}. Here, we apply it in a second-order nonlinear (Pockels) medium.

The paper is organized as follows. We derive the equations for the field operators of the pump and down-converted fields in Sec.~\ref{eom} and use them to obtain an expression for the infinitesimal propagation operator in Sec.~\ref{infpropop}. It is then used in Sec.~\ref{fipe} to derive a functional evolution equation. In Sec.~\ref{eenvoud}, we simplify this equation and in Sec.~\ref{oplos} we consider some of its solutions. Applications are discussed generically in Sec.~\ref{applic}. We end with conclusions in Sec.~\ref{concl}.

\section{\label{eom}Equations of motion}

\subsection{\label{metode}Infinitesimal propagation approach}

If $\hat{U}(z)$ represents that unitary evolution operator for the propagation of a state over a distance $z$ through a (nonlinear) medium and $\hat{\rho}$ is the density operator for the input state, then the output state is given by
\begin{equation}
\hat{\rho}(z) = \hat{U}(z)\hat{\rho}(0)\hat{U}^{\dag}(z) .
\label{evol}
\end{equation}
For an infinitesimal propagation distance $\Delta$, we have
\begin{equation}
\hat{U}(\Delta) \approx \mathds{1}-\frac{i\Delta}{\hbar} \hat{P}_{\Delta} ,
\end{equation}
where $\mathds{1}$ is the identity operator and $\hat{P}_{\Delta}$ represents the infinitesimal propagation. The negative sign comes from the phase convention: {\em phase increases with time}. Hence, it decreases with distance. In the limit $\Delta\rightarrow 0$, an {\em infinitesimal propagation equation} (IPE) is obtained:
\begin{equation}
i\hbar\frac{\text{d}}{\text{d}z} \hat{\rho}(z) = [\hat{P}_{\Delta},\hat{\rho}(z)] .
\label{ipe}
\end{equation}
It resembles the Heisenberg equation and the infinitesimal propagation operator $\hat{P}_{\Delta}$ is analogues to the Hamiltonian. However, here the evolution is with respect to propagation distance $z$ instead of time.

The current application of the infinitesimal propagation approach differs from the way it is used in the context of scintillation \cite{ipfe} in that full knowledge of the infinitesimal propagation operator for the nonlinear medium can be assumed, as opposed to statistical knowledge only for the scintillating medium. As a result, the current analysis does not need an ensemble average. The process remains completely unitary.

\subsection{Nonlinear wave equation}

To derive the evolution equation, we need to know the infinitesimal propagation operator. To obtain an expression for it, we start with the classical process.

When light propagates through a medium, the excitation in the medium behaves as a vector field and satisfies the usual wave equation for a linear medium. In a nonlinear medium the refractive index is modified by the strength of the vector field, producing additional nonlinear terms that are added to the linear equation. The leading nonlinear term contains two factors of the vector field contracted on a rank-three tensor representing the second-order nonlinear susceptibility of the medium.

In the analysis, we'll employ the paraxial approximation, which requires a monochromatic approximation. It allows us to represent the vector field as a phasor field. Thus, we can express the nonlinear wave equation as
\begin{align}
0 = & \nabla^2 \mathbf{E}(\mathbf{x},\omega_0) + n^2 k_0^2 \mathbf{E}(\mathbf{x},\omega_0) \nonumber \\
& +\frac{3}{2} k_0^2 \int_0^{\infty} \mathbf{E}(\mathbf{x},\omega)\cdot\chi\cdot\mathbf{E}(\mathbf{x},\omega_0-\omega) \nonumber \\
& +2\mathbf{E}(\mathbf{x},\omega)\cdot\chi\cdot\mathbf{E}^*(\mathbf{x},\omega-\omega_0)\ \frac{\text{d}\omega}{2\pi} .
\label{tussen0}
\end{align}
Here $\mathbf{E}(\mathbf{x},\omega_0)$ is the electric phasor field in the medium, $k_0=\omega_0/c$ is the vacuum wave number, $n=n(\omega_0)$ is the dispersive refractive index for a given polarization at the given frequency and $\chi$ is the nonlinear vertex rule. The latter is a rank-three susceptibility tensor with two electric phasor fields contracted on it.

To reduce the complexity in the analysis, we only consider type I phase-matching. It allows us to remove the indices by selecting appropriate polarization components and treat them as scalar fields. Other scenarios can be likewise investigated with only slight modifications.

The two terms under the integral in Eq.~(\ref{tussen0}) represent different processes. The first term represents a spontaneous parametric down-conversion process, while the last term represents difference frequency generation. They are distinguished by the fact the $\omega_0$ is either the pump frequency or one of the down-converted frequencies.

Considering evolution along a specific propagation direction (the $z$-direction), we separate the transverse direction from the longitudinal direction and express the transverse coordinates of the field in the Fourier domain. The result represents the fields as $z$-dependent spectral functions. For the paraxial approximation, we also pull out a $z$-directed plane wave from all fields. The remaining part of the field is slow varying in $z$.

These modifications are applied to the fields in Eq.~(\ref{tussen0}) and the second-order $z$-derivative is discarded. The result is the classical nonlinear paraxial wave equation appropriate for the conditions under investigation.

\subsection{Quantization of the electric field}

The expressions of the classical equations can be used as the basis for the development of the quantum theory. First, we convert the classical equations into equations for single-photon states, replacing the classical fields with quantized field operators.
The quantized field in a dielectric medium is given by
\begin{align}
\hat{\mathbf{E}}(\mathbf{x},t) = & \sum_{s} \int \sqrt{\frac{n\mu_0\hbar}{2}} \left[ -i \vec{\eta}_s \hat{a}_s(\mathbf{K},\omega)
\exp(i\omega t-i n \mathbf{k}\cdot\mathbf{x}) \right. \nonumber \\
& \left. + \text{h.c.} \right]\ \frac{\omega\ \text{d}\omega\ \text{d}^2 k}{(2\pi)^3 k_z} ,
\label{elekobv}
\end{align}
where $\vec{\eta}_s$ is the spin vector, with spin index $s$. The quantized field is expressed in terms of {\em optical beam variables}, which are defined in Appendix~\ref{obv}.

The expression for the quantized field operator is turned into a scalar field operator by computing the dot-product with a specific polarization vector. Fourier transforms are performed with respect to time and the transverse coordinates. The result is
\begin{equation}
\hat{E}^{(+)} \stackrel{\text{FT}}{\longrightarrow} -i\sqrt{\frac{n\hbar}{2\epsilon_0}}\ \frac{k_0}{k_{0z}} \hat{a}(\mathbf{K},\omega_0)
\exp(-i n k_{0z} z) .
\end{equation}
The classical fields already have a factor $\exp(-i n k_z z)$ removed. So, we do the same in the quantized field. At the same time, we assume that $k_{0z}\approx k_0$ under the paraxial approximation. Hence, we define
\begin{equation}
\hat{G}(\mathbf{K},\omega_0,z) \equiv -i \sqrt{\frac{n\hbar}{2\epsilon_0}}\ \hat{a}(\mathbf{K},\omega_0) ,
\label{gopdef}
\end{equation}
where the $z$-dependence is due to the nonlinearity of the medium. (A plane wave created at $z$ would generate other plane waves at other values of $z$ due to coupling.) The Hermitian adjoint gives the creation operator. Their equal-$z$ commutation relation is
\begin{align}
& [\hat{G}(\mathbf{K},\omega_0,z), \hat{G}^{\dag}(\mathbf{K}',\omega_0',z)] \nonumber \\
= & \frac{n\hbar}{2\epsilon_0} (2\pi)^3 k_{0z} \delta(\mathbf{K}-\mathbf{K}') \delta(\omega_0-\omega_0') ,
\label{commut0}
\end{align}
based on the commutation relation for the ladder operators in optical beam variables, given in Eq.~(\ref{commutmf}).

\subsection{Paraxial operator equations}

The classical scalar, paraxial, slow-varying spectral functions in the nonlinear paraxial wave equation are replaced by $\hat{G}(\mathbf{K},\omega_0,z)$ to obtain separate single-photon operator equations for the pump and down-converted fields in the form of nonlinear paraxial wave equations. They are given by
\begin{widetext}
\begin{align}
\begin{split}
i n_{\text{p}} k_{\text{p}} \partial_z \hat{G}_{\text{p}}(\mathbf{K},\omega_{\text{p}},z) = & -n_{\text{p}}^2 |\mathbf{K}|^2 \hat{G}_{\text{p}}(\mathbf{K},\omega_{\text{p}},z) + \frac{d_{\text{ooe}}k_{\text{p}}^2}{2\epsilon_0}
\int \hat{G}_{\text{d}}(\mathbf{K}_1,\omega_1,z) \hat{G}_{\text{d}}(\mathbf{K}_2,\omega_2,z)  \exp(i\Delta k_z z) \\
& \times (2\pi)^3 \delta(\omega_{\text{p}}-\omega_1-\omega_2)\delta(n_{\text{p}}\mathbf{K}-n_1\mathbf{K}_1-n_2\mathbf{K}_2)
n_1^2 n_2^2\ \dbar k_1\ \dbar k_2 , \\
i n_1 k_1 \partial_z \hat{G}_{\text{d}}(\mathbf{K}_1,\omega_1,z) = & -n_1^2 |\mathbf{K}_1|^2 \hat{G}_{\text{d}}(\mathbf{K}_1,\omega_1,z)
+ \frac{d_{\text{ooe}}k_1^2}{\epsilon_0}
\int \hat{G}_{\text{p}}(\mathbf{K}_0,\omega_0,z) \hat{G}_{\text{d}}^*(\mathbf{K}_2,\omega_2,z) \exp(-i\Delta k_z z) \\
& \times (2\pi)^3 \delta(\omega_0-\omega_1-\omega_2) \delta\left(n_0\mathbf{K}_0-n_1\mathbf{K}_1-n_2\mathbf{K}_2\right)
n_0^2 n_2^2\ \dbar k_0\ \dbar k_2 ,
\end{split}
\label{ipeabs}
\end{align}
\end{widetext}
where $\hat{G}_{\text{p}}(\mathbf{K},\omega,z)$ and $\hat{G}_{\text{d}}(\mathbf{K},\omega,z)$ are the annihilation operators for the pump and down-converted fields, respectively, $n_{\text{p}}\equiv n_{\text{eff}}(\omega_{\text{p}})$,
$n_1\equiv n_{\text{o}}(\omega_1)$, $n_2\equiv n_{\text{o}}(\omega-\omega_1)$,
\begin{equation}
\Delta k_z \equiv n_{\text{p}} k_{\text{p}z}-n_1 k_{1z}-n_2 k_{2z} ,
\label{dkz}
\end{equation}
the integration measure is defined in Eq.~(\ref{dk3nabv}), and
\begin{equation}
d_{\text{ooe}} \equiv 3\epsilon_0 \chi^{(2)}_{abc} \eta_a^{(\text{o})} \eta_b^{(\text{o})} \eta_c^{(\text{e})}
\label{defdooe}
\end{equation}
represents the strength of the nonlinear interaction for type I phase-matching.

The general expression for $\Delta k_z$ depends on the different frequencies and angles of the down-converted beams. If, we can assume that such angles are small enough to ignore, then we can use the expression for collinear, non-degenerate critical phase matching, given by
\begin{equation}
\Delta k_z = \frac{c n_1 n_2 |\mathbf{K}_1\omega_2-\mathbf{K}_2\omega_1|^2}{2\omega_1\omega_2\omega_{\text{p}} n_{\text{p}}} ,
\label{dkz0}
\end{equation}
where $\omega_1=\omega$ and $\omega_2=\omega_{\text{p}}-\omega$.

\section{\label{infpropop}Infinitesimal propagation operator}

In analogy to the Heisenberg equation, the IPE in Eq.~(\ref{ipe}) should apply to any operator, not only the density operator of a state, without any change to the infinitesimal propagation operator $\hat{P}_{\Delta}$. Therefore, one can also apply it to the field operators:
\begin{align}
\begin{split}
i\hbar\partial_z \hat{G}_{\text{p}}(\mathbf{K}',\omega',z)
= & \left[ \hat{P}_{\Delta}, \hat{G}_{\text{p}}(\mathbf{K}',\omega',z) \right] , \\
i\hbar\partial_z \hat{G}_{\text{d}}(\mathbf{K}',\omega',z)
= & \left[ \hat{P}_{\Delta}, \hat{G}_{\text{d}}(\mathbf{K}',\omega',z) \right] .
\end{split}
\label{ipepd}
\end{align}
The result should reproduce the equations in Eq.~(\ref{ipeabs}). We'll use this correspondence to obtain an expression for the infinitesimal propagation operator. For this purpose, we compute the commutations with the field operators, using Eq.~(\ref{commut0}).
The $z$-components of the wave vectors that comes from the measures are replaced with their wave numbers under the paraxial approximation. One can use the symmetries of the kernel functions to simplify the expressions. The commutations with the creation operators for the pump and down-converted fields, respectively, lead to the adjoint equations.

The expression for the proposed infinitesimal propagation operator has the form
\begin{align}
\hat{P}_{\Delta} = & \int H_{\text{p}}(\mathbf{K},\omega)\hat{G}_{\text{p}}^{\dag}(\mathbf{K},\omega,z)\hat{G}_{\text{p}}(\mathbf{K},\omega,z)\ \dbar k \nonumber \\
& + \int H_{\text{d}}(\mathbf{K},\omega) \hat{G}_{\text{d}}^{\dag}(\mathbf{K},\omega,z) \hat{G}_{\text{d}}(\mathbf{K},\omega,z)\ \dbar k \nonumber \\
& + \int V^*(\mathbf{K},\mathbf{K}_1,\mathbf{K}_2,\omega,\omega_1,\omega_2) \hat{G}_{\text{d}}^{\dag}(\mathbf{K}_1,\omega_1,z) \nonumber \\
& \times \hat{G}_{\text{d}}^{\dag}(\mathbf{K}_2,\omega_2,z) \hat{G}_{\text{p}}(\mathbf{K},\omega,z)\ \dbar k\ \dbar k_1\ \dbar k_2 \nonumber \\
& + \int V(\mathbf{K},\mathbf{K}_1,\mathbf{K}_2,\omega,\omega_1,\omega_2) \hat{G}_{\text{p}}^{\dag}(\mathbf{K},\omega,z) \nonumber \\
& \times \hat{G}_{\text{d}}(\mathbf{K}_1,\omega_1,z) \hat{G}_{\text{d}}(\mathbf{K}_2,\omega_2,z)\ \dbar k\ \dbar k_1\ \dbar k_2 ,
\label{pipo}
\end{align}
where $V(\mathbf{K},\mathbf{K}_1,\mathbf{K}_2,\omega,\omega_1,\omega_2)$ is an interaction vertex kernel for the nonlinear PDC process, and $H_{\text{p}}(\mathbf{K},\omega)$ and $H_{\text{d}}(\mathbf{K},\omega)$ are the kernel functions for the linear propagation of the pump and down-converted fields, respectively. The expressions of these kernel functions, as obtained by comparing Eq.~(\ref{ipeabs}) with the results after substituting Eq.~(\ref{pipo}) into the Eq.~(\ref{ipepd}), are
\begin{align}
\begin{split}
H_{\text{p}} = & \frac{2\epsilon_0}{k} |\mathbf{K}|^2 , \\
H_{\text{d}} = & \frac{2\epsilon_0}{k_1} |\mathbf{K}_1|^2 , \\
V = & - (2\pi)^3 d_{\text{ooe}}\frac{n_1^2 n_2^2}{n_0^2} k_1 k_2 k \exp(i\Delta k_z z) \\
& \times \delta(n_0\mathbf{K}-n_1\mathbf{K}_1-n_2\mathbf{K}_2) \delta(\omega-\omega_1-\omega_2) ,
\end{split}
\end{align}
where $k=\omega/c$ is the wave number at the pump frequency and $n_0\equiv n_{\text{eff}}(\omega)$ is the effective refractive index for the pump field.

\section{\label{fipe}Functional evolution equation}

For an investigation that incorporates all the degrees of freedom, it is convenient to consider the IPE in terms of Wigner functionals \cite{stquad,stquaderr,wigfunk}. The evolution equation in Eq.~(\ref{ipe}) can be carried over directly in terms of Wigner functionals where the commutation is expressed in terms of star products, because the quadrature basis elements that are required for the conversion of the operators to Wigner functionals do not depend on $z$. The variables of these functionals are functions or fields that are here denoted by lower case Greek symbols, such as $\alpha$ and $\alpha^*$. The functional variables are here referred to as {\em field variables}. They do not depend on $z$. The $z$-dependence is carried in the parameter functions for the state and the kernel function of the PDC vertex.

\subsection{\label{wiginfprop}Functional for the propagation operator}

To compute the Wigner functional for the infinitesimal propagation operator given in Eq.~(\ref{pipo}), we use a coherent state assisted approach, presented in Appendix~\ref{csap}. For this purpose, we need to determine what happens when the field operator, defined in Eq.~(\ref{gopdef}), is applied to a coherent state. It gives
\begin{align}
\hat{G}(\mathbf{K},\omega,z)\ket{\alpha}
= & -i \sqrt{\frac{n\hbar}{2\epsilon_0}}\ \hat{a}(\mathbf{K},\omega)\ket{\alpha} \nonumber \\
= & \ket{\alpha} \sqrt{\frac{n\hbar}{2\epsilon_0}}\ \alpha(\mathbf{K},\omega) ,
\label{gopvac}
\end{align}
where we absorb the $-i$ in the spectral function $\alpha(\mathbf{K},\omega)$. The relationship is valid for a fixed value of $z$, so that the $z$-dependence does not carry over to $\alpha$.

The overlap of the infinitesimal propagation operator by coherent states for the pump field and the down-converted fields on both sides gives
\begin{align}
& \bra{\alpha_1}\bra{\beta_1}\hat{P}_{\Delta}\ket{\alpha_2}\ket{\beta_2} \nonumber \\
= & \exp \left(\alpha_1^*\diamond\alpha_2+\beta_1^*\diamond\beta_2-\tfrac{1}{2}\|\alpha_1\|^2-\tfrac{1}{2}\|\alpha_2\|^2 \right. \nonumber \\
& \left. -\tfrac{1}{2}\|\beta_1\|^2-\tfrac{1}{2}\|\beta_2\|^2 \right)  \left(\beta_1^*\diamond P_{\text{p}}\diamond\beta_2+\alpha_1^*\diamond P_{\text{d}}\diamond\alpha_2 \right. \nonumber \\
& \left. -\beta_1^*\diamond T\diamond\diamond\ \alpha_2\alpha_2-\alpha_1^*\alpha_1^*\diamond\diamond\ T^*\diamond\beta_2\right) ,
\label{apa}
\end{align}
where $\beta_1$ and $\beta_2$ are associated with the pump field, $\alpha_1$ and $\alpha_2$ are associated with the down-converted field, and the $\diamond$-contraction is defined as \cite{wigfunk}
\begin{equation}
\alpha_1^*\diamond\alpha_2 \equiv \int \alpha_1^*(\mathbf{k}) \alpha_2(\mathbf{k})\ \dbar k .
\end{equation}
Furthermore,
\begin{align}
\begin{split}
P_{\text{p}} \equiv & \frac{n_0\hbar}{2\epsilon_0} H_{\text{p}} \mathbf{1}
= \frac{n_0\hbar}{k} |\mathbf{K}|^2 \mathbf{1} , \\
P_{\text{d}} \equiv & \frac{n_1\hbar}{2\epsilon_0} H_{\text{d}} \mathbf{1}
= \frac{n_1\hbar}{k_1} |\mathbf{K}_1|^2 \mathbf{1} , \\
T \equiv & - \sqrt{n_1 n_2 n_0}\left(\frac{\hbar}{2\epsilon_0}\right)^{3/2} V \\
= & (2\pi)^3 \frac{\hbar \sigma_{\text{ooe}}}{c^{7/2}} \omega_1 \omega_2 \omega \exp(i \Delta k_z z) \\
& \times \delta(n_0\mathbf{K}-n_1\mathbf{K}_1-n_2\mathbf{K}_2) \delta(\omega-\omega_1-\omega_2) ,
\end{split}
\label{kernel0}
\end{align}
where
\begin{equation}
\mathbf{1} \equiv (2\pi)^3 k_{z} \delta(\mathbf{K}-\mathbf{K}') \delta(\omega-\omega') ,
\label{eendef}
\end{equation}
and the cross-section (having the units of an area) for the PDC process is defined by
\begin{equation}
\sigma_{\text{ooe}} = \frac{3}{2} \sqrt{\frac{c\hbar}{2\epsilon_0} \frac{(n_1 n_2)^5}{n_0^3}}
\chi^{(2)}_{abc} \eta_a^{(\text{o})} \eta_b^{(\text{o})} \eta_c^{(\text{e})} .
\label{sigmadef}
\end{equation}

To complete the calculation of the Wigner functional, we substitute Eq.~(\ref{apa}) into the coherent state assisted functional integral expression in Eq.~(\ref{asscohwig}) and evaluate the integrals. The integration process can be alleviated by using a generating functional together with a construction operator. The generating functional is
\begin{align}
\mathcal{G} = & \exp \left(\alpha_1^*\diamond\alpha_2+\beta_1^*\diamond\beta_2-\tfrac{1}{2}\|\alpha_1\|^2-\tfrac{1}{2}\|\alpha_2\|^2
\right. \nonumber \\
& \left. -\tfrac{1}{2}\|\beta_1\|^2-\tfrac{1}{2}\|\beta_2\|^2 + \alpha_1^*\diamond\mu_1 + \mu_2^*\diamond\alpha_2 \right. \nonumber \\
& \left. + \beta_1^*\diamond\eta_1 + \eta_2^*\diamond\beta_2 \right) ,
\end{align}
where $\mu_1$, $\mu_2^*$, $\eta_1$, and $\eta_2^*$ are auxiliary fields. The construction operator is defined using functional derivatives
\begin{align}
\hat{\mathcal{C}} =
& - \frac{\delta}{\delta\eta_1} \diamond T\diamond\diamond\ \frac{\delta}{\delta\mu_2^*} \frac{\delta}{\delta\mu_2^*}
- \frac{\delta}{\delta\mu_1} \frac{\delta}{\delta\mu_1} \diamond\diamond\ T^* \diamond \frac{\delta}{\delta\eta_2^*} \nonumber \\
& + \frac{\delta}{\delta\eta_1} \diamond P_{\text{p}} \diamond \frac{\delta}{\delta\eta_2^*}
+ \frac{\delta}{\delta\mu_1} \diamond P_{\text{d}} \diamond \frac{\delta}{\delta\mu_2^*} .
\label{konstruk}
\end{align}
When the generating functional is substituted into the coherent state assisted functional integral expression in Eq.~(\ref{asscohwig}) and all the functional integrals are evaluated, we obtain a generating functional given by
\begin{align}
\mathcal{W} = & \exp\left(\mu_2^*\diamond\alpha+\alpha^*\diamond\mu_1+\eta_2^*\diamond\beta+\beta^*\diamond\eta_1 \right. \nonumber \\
& \left. -\tfrac{1}{2}\eta_2^*\diamond\eta_1-\tfrac{1}{2}\mu_2^*\diamond\mu_1 \right) .
  \label{wigpropgen}
\end{align}
To obtain the Wigner functional for the infinitesimal propagation operator, one would apply the construction operator and set the sources to zero. However, it is not convenient to use the expression in this form.

\subsection{\label{wigipe}Infinitesimal propagation equation}

Instead of using the Wigner functional for the infinitesimal propagation operator directly in the expression of the IPE, we find it more convenient to represent it in terms of the construction operator in Eq.~(\ref{konstruk}) and the generating functional in Eq.~(\ref{wigpropgen}). The IPE in terms of Wigner functionals then reads
\begin{equation}
i\hbar \frac{\text{d}}{\text{d}z} W_{\hat{\rho}} = \left. \hat{\mathcal{C}}\left\{ \mathcal{W}\star W_{\hat{\rho}}-W_{\hat{\rho}}\star\mathcal{W} \right\} \right|_{\{\mu,nu\}=0} ,
\label{ipewig0}
\end{equation}
where $W_{\hat{\rho}}[\alpha^*,\alpha,\beta^*,\beta]$ is the Wigner functional for the complete photonic state (pump and down-converted field), with $\alpha$ and $\beta$ being the field variables associated with the down-converted field and the pump field, respectively, and $\star$ represents the star product.

We evaluate the functional integrals for the two star products in turn. Two of the functional integrals in each produce Dirac-$\delta$ functionals, which are removed by the remaining two functional integrals. As a result, we obtain
\begin{align}
& \mathcal{W}\star W_{\hat{\rho}}[\alpha^*,\alpha,\beta^*,\beta] - W_{\hat{\rho}}[\alpha^*,\alpha,\beta^*,\beta]\star\mathcal{W} \nonumber \\
= & W_{\hat{\rho}}\left[\alpha^*+\frac{\mu_2^*}{2},\alpha-\frac{\mu_1}{2},
\beta^*+\frac{\eta_2^*}{2},\beta-\frac{\eta_1}{2}\right] \mathcal{W} \nonumber \\
& - W_{\hat{\rho}}\left[\alpha^*-\frac{\mu_2^*}{2},\alpha+\frac{\mu_1}{2},
\beta^*-\frac{\eta_2^*}{2},\beta+\frac{\eta_1}{2}\right] \mathcal{W} .
\label{sterprod}
\end{align}
We can now substitute these two terms into Eq.~(\ref{ipewig0}), apply the construction operator and set the source fields to zero. When the result is expressed in compact notation, the double contractions can be ambiguous. Therefore, we express the evolution equation in terms of integrals over the three-dimensional wave vectors:
\begin{widetext}
\begin{align}
i\hbar\frac{\text{d} W_{\hat{\rho}}}{\text{d}z} = & \int \left[ \beta^*(\mathbf{k}_1) P_{\text{p}}(\mathbf{k}_1,\mathbf{k}_2)\frac{\delta W_{\hat{\rho}}}{\delta\beta^*(\mathbf{k}_2)}
-\frac{\delta W_{\hat{\rho}}}{\delta\beta(\mathbf{k}_1)}P_{\text{p}}(\mathbf{k}_1,\mathbf{k}_2)\beta(\mathbf{k}_2) \right. \nonumber \\
& \left. +\alpha^*(\mathbf{k}_1)P_{\text{d}}(\mathbf{k}_1,\mathbf{k}_2)\frac{\delta W_{\hat{\rho}}}{\delta\alpha^*(\mathbf{k}_2)}
-\frac{\delta W_{\hat{\rho}}}{\delta\alpha(\mathbf{k}_1)}P_{\text{d}}(\mathbf{k}_1,\mathbf{k}_2)\alpha(\mathbf{k}_2) \right]\ \dbar k_1\ \dbar k_2\nonumber \\
& + \int \left[ \frac{\delta W_{\hat{\rho}}}{\delta\beta(\mathbf{k}_3)}T(\mathbf{k}_1,\mathbf{k}_2,\mathbf{k}_3,z) \alpha(\mathbf{k}_1) \alpha(\mathbf{k}_2)
- 2\beta^*(\mathbf{k}_3)T(\mathbf{k}_1,\mathbf{k}_2,\mathbf{k}_3,z) \alpha(\mathbf{k}_1) \frac{\delta W_{\hat{\rho}}}{\delta\alpha^*(\mathbf{k}_2)} \right. \nonumber \\
& -\alpha^*(\mathbf{k}_1) \alpha^*(\mathbf{k}_2)T^*(\mathbf{k}_1,\mathbf{k}_2,\mathbf{k}_3,z)\frac{\delta W_{\hat{\rho}}}{\delta\beta^*(\mathbf{k}_3)}
+2\frac{\delta W_{\hat{\rho}}}{\delta\alpha(\mathbf{k}_1)} \alpha^*(\mathbf{k}_2)T^*(\mathbf{k}_1,\mathbf{k}_2,\mathbf{k}_3,z)\beta(\mathbf{k}_3)\nonumber \\
& \left. +\frac{1}{4} T(\mathbf{k}_1,\mathbf{k}_2,\mathbf{k}_3,z)
\frac{\delta^3 W_{\hat{\rho}}}{\delta\beta(\mathbf{k}_3)\delta\alpha^*(\mathbf{k}_1)\delta\alpha^*(\mathbf{k}_2)}
-\frac{1}{4} T^*(\mathbf{k}_1,\mathbf{k}_2,\mathbf{k}_3,z)
\frac{\delta^3 W_{\hat{\rho}}}{\delta\alpha(\mathbf{k}_1)\delta\alpha(\mathbf{k}_2)\delta\beta^*(\mathbf{k}_3)} \right]\ \dbar k_1\ \dbar k_2\ \dbar k_3 .
\label{ipewig1}
\end{align}
\end{widetext}

The functional differential equation in Eq.~(\ref{ipewig1}) is the evolution equation for the complete state during PDC. It is valid for all possible scenarios and can handle situations where the pump is any kind of state, which may become entangled with the down-converted state during the PDC process. However, the complicated expression of the evolution equation makes it challenging to find a general solution for the state.

Since it is linear in the Wigner functional of the state, the solution would be represented in exponential form with a polynomial functional exponent. Unfortunately, as shown below, the terms in the polynomial functional does not close. At every order, the equation generates higher order terms. Therefore, a general solution would be represented as an exponential function with a polynomial functional of infinite order in its argument.

\section{\label{eenvoud}Simplifications}

Having obtained the general equation for the state produced by the PDC process, we are presented by the challenge to solve it. First, we consider some simplifications and approximations that would allow one to find solutions under certain conditions.

\subsection{\label{refframe}Reference frame representation}

The first simplification is introduced to reduce the number of terms in the equation, by removing those terms that contain $P_{\text{p}}$ and $P_{\text{d}}$. Formally, the field variables are transformed as
\begin{align}
\begin{split}
\alpha \equiv & U_{\text{d}}(z)\diamond\bar{\alpha} , \\
\alpha^* \equiv & \bar{\alpha}^*\diamond U_{\text{d}}^{\dag}(z) , \\
\beta \equiv & U_{\text{p}}(z)\diamond\bar{\beta} , \\
\beta^* \equiv & \bar{\beta}^*\diamond U_{\text{p}}^{\dag}(z) ,
\end{split}
\label{cotra}
\end{align}
where the barred field variables represent new field variables in terms of which the equation will be expressed and $U_{\text{d}}(z)$ and $U_{\text{p}}(z)$ represent unitary kernels for linear propagation of the pump and down-converted fields along $z$, respectively. The transformation maps the field space back onto itself. Therefore, the fields won't change, but the points where they are applied will change.

We replace the arguments of the Wigner functional with the transformed fields:
\begin{align}
W_{\hat{\rho}}[\alpha^*,\alpha,\beta^*,\beta](z)
= & W_{\hat{\rho}}\left[ \bar{\alpha}^*\diamond U_{\text{d}}^{\dag}(z), U_{\text{d}}(z)\diamond\bar{\alpha}, \right. \nonumber \\
& \left. \bar{\beta}^*\diamond U_{\text{p}}^{\dag}(z), U_{\text{p}}(z)\diamond\bar{\beta} \right](z) \nonumber \\
\equiv &\ \widetilde{W}_{\hat{\rho}}[\bar{\alpha}^*,\bar{\alpha},\bar{\beta}^*,\bar{\beta}](z) .
\label{deftildew}
\end{align}
Then we apply the total $z$-derivative, leading to
\begin{align}
\frac{\text{d} W_{\hat{\rho}}}{\text{d}z} = & \bar{\alpha}^*\diamond \partial_z U_{\text{d}}^{\dag}(z)\diamond\frac{\delta W_{\hat{\rho}}}{\delta\alpha^*}
+ \frac{\delta W_{\hat{\rho}}}{\delta\alpha}\diamond \partial_z U_{\text{d}}(z)\diamond\bar{\alpha} \nonumber \\
& + \bar{\beta}^*\diamond \partial_z U_{\text{p}}^{\dag}(z)\diamond\frac{\delta W_{\hat{\rho}}}{\delta\beta^*}
+ \frac{\delta W_{\hat{\rho}}}{\delta\beta}\diamond \partial_z U_{\text{p}}(z)\diamond\bar{\beta} \nonumber \\
& + \partial_z W_{\hat{\rho}} .
\end{align}
A comparison with Eq.~(\ref{ipewig1}) indicates that, if
\begin{align}
\begin{split}
U_{\text{p}}(z) \equiv & \exp_{\diamond}\left(\frac{i}{\hbar} z P_{\text{p}}\right)
= \exp\left(\frac{i z c n_0}{\omega} |\mathbf{K}|^2 \right) \mathbf{1} , \\
U_{\text{d}}(z) \equiv & \exp_{\diamond}\left(\frac{i}{\hbar} z P_{\text{d}}\right)
= \exp\left(\frac{i z c n_1}{\omega_1} |\mathbf{K}_1|^2 \right) \mathbf{1} ,
\end{split}
\label{defuus}
\end{align}
then the total derivative with respect to $z$ would produce the first four terms in Eq.~(\ref{ipewig1}). The subscript $\diamond$ indicates a functional whose expansion has a first term given by $\mathbf{1}$ and all products are represented by $\diamond$-contractions.

The transformation also implies that the functional derivatives become
\begin{align}
\begin{split}
\frac{\delta W_{\hat{\rho}}}{\delta\alpha^*}
= & \frac{\delta\widetilde{W}_{\hat{\rho}}}{\delta\bar{\alpha}^*}\diamond U_{\text{d}}(z) , \\
\frac{\delta W_{\hat{\rho}}}{\delta\alpha}
= & U_{\text{d}}^{\dag}(z)\diamond\frac{\delta \widetilde{W}_{\hat{\rho}}}{\delta\bar{\alpha}} , \\
\frac{\delta W_{\hat{\rho}}}{\delta\beta^*}
= & \frac{\delta \widetilde{W}_{\hat{\rho}}}{\delta\bar{\beta}^*}\diamond U_{\text{p}}(z) , \\
\frac{\delta W_{\hat{\rho}}}{\delta\beta}
= & U_{\text{p}}^{\dag}(z)\diamond\frac{\delta \widetilde{W}_{\hat{\rho}}}{\delta\bar{\beta}} .
\end{split}
\end{align}
The additional factors of the unitary propagation kernels ($U_{\text{p}}$, $U_{\text{d}}$ and their Hermitian adjoints) are now attached to the vertex kernel. Therefore, the vertex kernel is transformed as follows:
\begin{align}
T(z) \rightarrow & \int U_{\text{d}}(\mathbf{k}_1,\mathbf{k}_1',z) U_{\text{d}}(\mathbf{k}_2,\mathbf{k}_2',z) T(\mathbf{k}_1',\mathbf{k}_2',\mathbf{k}',z) \nonumber \\
& \times U_{\text{p}}^{\dag}(\mathbf{k}',\mathbf{k},z)\ \dbar k_1'\ \dbar k_2'\ \dbar k' \nonumber \\
\equiv &\ \widetilde{T}(\mathbf{k}_1,\mathbf{k}_2,\mathbf{k},z) .
\label{wyetee}
\end{align}

Applying the transformation to both the fields and the functional derivatives in the evolution equation, we get
\begin{align}
i\hbar\partial_z \widetilde{W}_{\hat{\rho}} = & \int \left[ \frac{\delta \widetilde{W}_{\hat{\rho}}}{\delta\bar{\beta}(\mathbf{k}_3)}\widetilde{T}(\mathbf{k}_1,\mathbf{k}_2,\mathbf{k}_3,z) \bar{\alpha}(\mathbf{k}_1) \bar{\alpha}(\mathbf{k}_2) \right. \nonumber \\
& - 2\bar{\beta}^*(\mathbf{k}_3)\widetilde{T}(\mathbf{k}_1,\mathbf{k}_2,\mathbf{k}_3,z) \bar{\alpha}(\mathbf{k}_1)
\frac{\delta \widetilde{W}_{\hat{\rho}}}{\delta\bar{\alpha}^*(\mathbf{k}_2)} \nonumber \\
& +\frac{1}{4} \widetilde{T}(\mathbf{k}_1,\mathbf{k}_2,\mathbf{k}_3,z)
\frac{\delta^3 \widetilde{W}_{\hat{\rho}}}{\delta\bar{\beta}(\mathbf{k}_3)\delta\bar{\alpha}^*(\mathbf{k}_1)\delta\bar{\alpha}^*(\mathbf{k}_2)} \nonumber \\
& -\bar{\alpha}^*(\mathbf{k}_1) \bar{\alpha}^*(\mathbf{k}_2)\widetilde{T}^*(\mathbf{k}_1,\mathbf{k}_2,\mathbf{k}_3,z)\frac{\delta \widetilde{W}_{\hat{\rho}}}{\delta\bar{\beta}^*(\mathbf{k}_3)}\nonumber \\
& +2\frac{\delta \widetilde{W}_{\hat{\rho}}}{\delta\bar{\alpha}(\mathbf{k}_1)} \bar{\alpha}^*(\mathbf{k}_2)\widetilde{T}^*(\mathbf{k}_1,\mathbf{k}_2,\mathbf{k}_3,z)\bar{\beta}(\mathbf{k}_3)\nonumber \\
& \left. -\frac{1}{4} \widetilde{T}^*(\mathbf{k}_1,\mathbf{k}_2,\mathbf{k}_3,z)
\frac{\delta^3 \widetilde{W}_{\hat{\rho}}}{\delta\bar{\alpha}(\mathbf{k}_1)\delta\bar{\alpha}(\mathbf{k}_2)\delta\bar{\beta}^*(\mathbf{k}_3)} \right]\ \nonumber \\
& \times \dbar k_1\ \dbar k_2\ \dbar k_3 .
\label{ipewigco}
\end{align}
Since the transformation maps the functional space onto itself, we can revert to the unbarred $\alpha$'s and $\beta$'s. On the other hand, we retain the tildes on $T$ and $W$ to indicate that they are ``dressed'' by the propagation kernels.

\subsection{\label{eksform}Exponential form}

The fact that Eq.~(\ref{ipewigco}) only contains terms that are linear in the Wigner functional of the state implies solutions that are in exponential form:
\begin{equation}
\widetilde{W}_{\hat{\rho}}[\alpha^*,\alpha,\beta^*,\beta](z) = \exp\left\{F[\alpha^*,\alpha,\beta^*,\beta](z)\right\} .
\label{ekswig}
\end{equation}
Here $F[\alpha^*,\alpha,\beta^*,\beta](z)$ is a multivariate polynomial functional, with coefficients in the form of $z$-dependent kernels. The equation for $F$ would contain more terms due to the third-order functional derivatives.

\subsection{\label{kohpomp}Coherent state pump}

So far, the simplifications were not restrictive in any way. They retained the full validity of the original equation. Here, we introduce an approximation that is only valid for certain experimental conditions.

In most experimental scenarios, the pump that illuminates the nonlinear crystal is considered to be a coherent state. Here, we assume that it remains a coherent state during the PDC process. It is only the pump's parameter function $\zeta(z)$ that changes as a function of $z$. However, since the squared magnitude of the parameter function $\|\zeta(z)\|^2$ represents the pump power, its evolution includes the possible depletion of the pump power and thus goes beyond the conditions for the semi-classical approximation. Therefore, we substitute
\begin{equation}
\widetilde{W}_{\hat{\rho}}[\alpha,\beta] \rightarrow \mathcal{N}_0 \exp\left[-2\|\beta-\zeta(z)\|^2\right] W_{\hat{\sigma}}[\alpha] ,
\label{pompans}
\end{equation}
into Eq.~(\ref{ipewigco}), where $W_{\hat{\sigma}}$ is the Wigner functional for the down-converted part only and evaluate the functional derivatives with respect to $\beta$. Then, the pump field is factored out and removed, leading to
\begin{widetext}
\begin{align}
i\hbar\partial_z W_{\hat{\sigma}} = & 2 \int \left\{ \alpha^*(\mathbf{k}_1) \alpha^*(\mathbf{k}_2) \widetilde{T}^*(\mathbf{k}_1,\mathbf{k}_2,\mathbf{k}_3,z) \left[\beta(\mathbf{k}_3)-\zeta(\mathbf{k}_3,z)\right] W_{\hat{\sigma}}
+ \left(\delta_1 W_{\hat{\sigma}}\right) \alpha^*(\mathbf{k}_2)
\widetilde{T}^*(\mathbf{k}_1,\mathbf{k}_2,\mathbf{k}_3,z) \beta(\mathbf{k}_3) \right. \nonumber \\
& +\frac{1}{4} \left(\delta_1 \delta_2 W_{\hat{\sigma}}\right) \widetilde{T}^*(\mathbf{k}_1,\mathbf{k}_2,\mathbf{k}_3,z)
\left[\beta(\mathbf{k}_3)-\zeta(\mathbf{k}_3,z)\right]
-\frac{1}{4} \left[\beta^*(\mathbf{k}_3)-\zeta^*(\mathbf{k}_3,z)\right] \widetilde{T}(\mathbf{k}_1,\mathbf{k}_2,\mathbf{k}_3,z)
\left(\delta_1^* \delta_2^* W_{\hat{\sigma}}\right) \nonumber \\
& \left. -\left[\beta^*(\mathbf{k}_3)-\zeta^*(\mathbf{k}_3,z)\right] \widetilde{T}(\mathbf{k}_1,\mathbf{k}_2,\mathbf{k}_3,z) \alpha(\mathbf{k}_1) \alpha(\mathbf{k}_2) W_{\hat{\sigma}}
-\beta^*(\mathbf{k}_3) \widetilde{T}(\mathbf{k}_1,\mathbf{k}_2,\mathbf{k}_3,z)
\alpha(\mathbf{k}_1) \left(\delta_2^* W_{\hat{\sigma}}\right) \right\}\ \dbar k_1\ \dbar k_2\ \dbar k_3 \nonumber \\
& -i 2\hbar\int \left\{ \left[\beta^*(\mathbf{k})-\zeta^*(\mathbf{k},z)\right]\partial_z\zeta(\mathbf{k},z)
+\partial_z\zeta^*(\mathbf{k},z)\left[\beta(\mathbf{k})-\zeta(\mathbf{k},z)\right] \right\}\ \dbar k\ W_{\hat{\sigma}} ,
\label{ipewig2}
\end{align}
\end{widetext}
where
\begin{equation}
\delta_n \equiv \frac{\delta}{\delta\alpha(\mathbf{k}_n)} ~~~~~
\delta_n^* \equiv \frac{\delta}{\delta\alpha^*(\mathbf{k}_n)} .
\end{equation}
Equation~(\ref{ipewig2}) represents a general expression for the evolution of the down-converted state during PDC.

\section{\label{oplos}Solutions}

Using the simpler equation in Eq.~(\ref{ipewig2}), we can now consider solutions. First, we reproduce the familiar semi-classical solution. Then we proceed to provide solutions beyond the semi-classical approximation. It includes a solution for an evolving coherent pump state and eventually for the down-converted state with the aid of a perturbative approach.

\subsection{Semi-classical approximation}

The equation in Eq.~(\ref{ipewig2}) is still rather complicated to solve in general. To simplify it further, we argue that, since $\beta$ can represent any field, we can set it equal to the parameter function $\zeta(z)$. As such, it samples only one point on the phase space of the pump, namely the point at the peak of the coherent state's Wigner functional. The substitution effectively converts the pump field into a classical field, and therefore implies a semi-classical approximation. The resulting equation simplifies significantly:
\begin{align}
i\hbar\partial_z W_{\hat{\sigma}} = & 2 \int \left[ \left(\delta_1 W_{\hat{\sigma}}\right)
\alpha^*(\mathbf{k}_2) \widetilde{T}^*(\mathbf{k}_1,\mathbf{k}_2,\mathbf{k}_3,z) \zeta(\mathbf{k}_3,z) \right. \nonumber \\
& \left. -\zeta^*(\mathbf{k}_3,z) \widetilde{T}(\mathbf{k}_1,\mathbf{k}_2,\mathbf{k}_3,z) \alpha(\mathbf{k}_1)
\left(\delta_2^* W_{\hat{\sigma}}\right) \right] \nonumber \\
& \times \dbar k_1\ \dbar k_2\ \dbar k_3 .
\label{semklas0}
\end{align}

Since the derivative terms for the parameter function of the pump dropped away, the parameter function does not evolve, apart from linear propagation. This situation reproduces the notion of an undepleted pump field, associated with the semi-classical approximation.

To simplify the notation, we incorporate the pump parameter function with the vertex kernel into a bilinear kernel function, which is defined by
\begin{equation}
H(\mathbf{k}_1,\mathbf{k}_2,z) = \frac{i4}{\hbar} \int \widetilde{T}(\mathbf{k}_1,\mathbf{k}_2,\mathbf{k},z) \zeta^*(\mathbf{k})\ \dbar k .
\label{defhaa}
\end{equation}
It allows us to revert to the $\diamond$-contraction notation.

We now use the following ansatz, which has the form of the Wigner functional for a squeezed vacuum state
\begin{align}
W_{\hat{\sigma}} = & \mathcal{N} \exp\left[-2\alpha^*\diamond A(z)\diamond\alpha \right. \nonumber \\
& \left. -\alpha\diamond B(z)\diamond\alpha-\alpha^*\diamond B^*(z)\diamond\alpha^*\right] ,
\label{ansatz}
\end{align}
where $\mathcal{N}$ is a normalization constant, which remains constant with $z$, and $A(z)$ and $B(z)$ are unknown kernel functions to be determined. The kernel function $A(z)$ is Hermitian and positive definite, and $B(z)$ is symmetric. After substituting it into Eq.~(\ref{semklas0}) and dividing by $i\hbar W_{\hat{\sigma}}$, we obtain an equation that can be separated into three equations. By removing the different combinations of $\alpha$ and $\alpha^*$ with the aid of functional derivatives, we obtain
\begin{align}
\begin{split}
\partial_z A(z) = & \tfrac{1}{2}H^*(z)\diamond B(z) + \tfrac{1}{2}B^*(z)\diamond H(z) , \\
\partial_z B(z) = & \tfrac{1}{2}H(z)\diamond A(z) + \tfrac{1}{2}A^T(z)\diamond H(z) , \\
\partial_z B^*(z) = & \tfrac{1}{2}A(z)\diamond H^*(z) + \tfrac{1}{2}H^*(z)\diamond A^T(z) ,
\end{split}
\label{vgees1}
\end{align}
where the transpose indicates that the two wave vectors in the argument of $A$ are interchanged.
To solve the equations in Eq.~(\ref{vgees1}), one can follow various approaches. One way is to integrate all the equations with respect to $z$ and then perform progressive back substitutions to obtain expansions in terms of integrals of contracted $H$-kernels. The initial conditions for the expansions are taken as $A(0)=\mathbf{1}$ and $B(0)=0$ for the Wigner functional of the vacuum state. The resulting expressions, which satisfy the equations in Eq.~(\ref{vgees1}) are
\begin{widetext}
\begin{align}
\begin{split}
A(z) = & \mathbf{1} + \int_0^z \int_0^{z_1} \mathcal{Z}\left\{ H^*(z_1)\diamond H(z_2) \right\}\ \text{d}z_2\ \text{d}z_1 \\
& + \int_0^z \int_0^{z_1} \int_0^{z_2} \int_0^{z_3} \mathcal{Z}\left\{ H^*(z_1)\diamond H(z_2)\diamond H^*(z_3)\diamond H(z_4)
\right\}\ \text{d}z_4\ \text{d}z_3\ \text{d}z_2\ \text{d}z_1 + ... , \\
B(z) = & \int_0^{z} H(z_1)\ \text{d}z_1 + \int_0^z \int_0^{z_1} \int_0^{z_2}
\mathcal{Z}\left\{ H(z_1)\diamond H^*(z_2)\diamond H(z_3) \right\}\ \text{d}z_3\ \text{d}z_2\ \text{d}z_1 \\
& + \int_0^z  \int_0^{z_1} \int_0^{z_2} \int_0^{z_3} \int_0^{z_4}
\mathcal{Z}\left\{ H(z_1)\diamond H^*(z_2)\diamond H(z_3) \diamond H^*(z_4) \diamond H(z_5) \right\} \\
& \times\ \text{d}z_5\ \text{d}z_4\ \text{d}z_3\ \text{d}z_2\ \text{d}z_1 + ...\ ,
\end{split}
\label{defzab}
\end{align}
where $\mathcal{Z}\{\cdot\}$ represents a $z$-symmetrization operation, recursively defined by
\begin{equation}
\mathcal{Z}\left\{ f_1(z_1)\diamond ...\diamond f_n(z_n) \right\} =
\frac{1}{2}\left[f_1(z_1)\diamond \mathcal{Z}\left\{ f_2(z_2)\diamond ...\diamond f_n(z_n) \right\}
+\mathcal{Z}\left\{ f_1(z_2)\diamond ...\diamond f_{n-1}(z_n) \right\}\diamond f_n(z_1) \right],
\end{equation}
with $\mathcal{Z}\{f_1(z_1)\}=f_1(z_1)$.
\end{widetext}

\subsection{\label{ander}Bloch-Messiah reduction}

The approach we followed to solve the set of equations in Eq.~(\ref{vgees1}) differs from the more general approach that is often used for PDC. The more general approach is to perform a Bloch-Messiah reduction \cite{blochmessiah,braunstein} that results in Bogoliubov transformations. Being based on linear algebra, the Bloch-Messiah reduction assumes a finite dimensional system. The kernels are then represented by matrices, which can be diagonalized, reminiscent of a Schmidt decomposition \cite{peres,thapliyal,sharapova,namiki}. Thus, the set of equations in Eq.~(\ref{vgees1}) becomes a decoupled set of ordinary differential equations that can be solved.

To address the case where the kernels are functions instead of matrices, the Bloch-Messiah reduction needs to have a well-defined continuous limit. Hence, the bilinear kernel must be represented by an infinite dimensional Schmidt-like decomposition:
\begin{equation}
H(\mathbf{k}_1,\mathbf{k}_2,z) = \sum_{m=0}^{\infty} h_m(z) \Phi_m(\mathbf{k}_1,z) \Phi_m(\mathbf{k}_2,z) ,
\end{equation}
where $\Phi_m(\mathbf{k},z)$ represents the Schmidt basis. Such a decomposition is represented by Mercer's theorem \cite{mercer}. For explicit calculations, one usually needs to determine the expressions for the Schmidt basis, which is a challenging endeavour. Therefore, we do not use the Bloch-Messiah reduction approach here.

The Bloch-Messiah reduction approach often produces semi-classical kernel functions in the form of hyperbolic trigonometric functions \cite{sharapova}. Here, such a result can be reproduced, in the {\em thin crystal limit}, where the Rayleigh range of the pump beam is much longer than the length of the nonlinear crystal, which is valid in most experimental scenarios. The expession of $H$ then becomes independent of $z$. Evaluating all the $z$-integrals and summing the terms, we obtain
\begin{align}
\begin{split}
A(z) \approx &\ \cosh_{\diamond} (z H_0) , \\
B(z) \approx &\ i \sinh_{\diamond} (z H_0) ,
\end{split}
\label{defzab0}
\end{align}
where we defined $H(z=0)\equiv i H_0$. The subscript $\diamond$ indicates that these kernels incorporate all the spatiotemporal degrees of freedom.

\subsection{\label{kohpompa}Coherent pump approximation}

Considering the more general case in Eq.~(\ref{ipewig2}), one can assess which terms are necessary in the polynomial functional $F$. If cross-terms (those that contain both $\alpha$'s and $\beta$'s) are needed, entanglement between the pump and the down-converted state may be inevitable. On the other hand, if such cross-terms can be shown to cancel among themselves, the state would be separable without any entanglement between the pump field and the down-converted field. Since $\alpha$ and $\beta$ are field variables, each combination of them represents an independent equation that would become zero in the separable case. For this purpose, we consider only those cross-terms in Eq.~(\ref{ipewig2}) that carry a factor of $\beta^*$. At the very least, the polynomial functional should contain the terms given in the ansatz in Eq.~(\ref{ansatz}). When we substitute them into those terms in Eq.~(\ref{ipewig2}) with a factor of $\beta^*$ and extract separate equations for the different combinations of $\alpha$ and $\alpha^*$, we obtain a set of four equations. One of these equations
\begin{align}
0 = & \int \widetilde{T}(\mathbf{k}_1,\mathbf{k}_2,\mathbf{k},z) B^*(\mathbf{k}_1,\mathbf{k}_3,z)\alpha^*(\mathbf{k}_3) \nonumber \\
& \times B^*(\mathbf{k}_2,\mathbf{k}_4,z)\alpha^*(\mathbf{k}_4)\ \dbar k_1\ \dbar k_2\ \dbar k_3\ \dbar k_4 ,
\label{vgees2d}
\end{align}
indicates that the required cancellations cannot work, because we do not expect $B^*$ to be zero.

We conclude that the pump and the down-converted fields are inevitably coupled in the description of the problem, which may imply that they are entangled by the PDC process. (Such entanglement between the pump and down-converted field has been theoretically predicted \cite{thopo} and experimentally observed \cite{eksopo1,eksopo2} in optical parametric oscillators.) Moreover, the polynomial functional need to be of infinite order, because at each order, higher order terms are generated in the equation. Such a situation does not allow us to obtain an exact closed form analytical solution for the equation. As a result, we are forced to use approximations to simplify the problem so that we can find a solution.

Another one of these equations
\begin{equation}
0 = -i\hbar\partial_z\zeta(z) + \frac{1}{2} \tr\left\{\widetilde{T}(z')\diamond B^*(z') \right\} ,
\label{vgees2a}
\end{equation}
where the trace only involves the down-converted fields' wave vectors, is an evolution equation for the pump parameter function. The solution is
\begin{equation}
\zeta(z) = \zeta(0) + \frac{1}{i 2\hbar} \int_0^z \tr\left\{\widetilde{T}(z')\diamond B^*(z') \right\}\ \text{d}z' .
\label{pompevol}
\end{equation}

Since $H(z)$, given in Eq.~(\ref{defhaa}), contains $\zeta(z)$ and $B^*(z)$ is defined in terms of $H(z)$, it appears that the equation for $\zeta(z)$ in Eq.~(\ref{pompevol}) contains multiple factors of $\zeta(z)$. The resulting equation would therefore be very difficult to solve. However, since the second term on the right-hand side of Eq.~(\ref{pompevol}) contains an unenhanced vertex, it is suppressed relative to the first term. An enhancement needs an extra factor of the parameter field, as found in $H(z)$. Therefore, up to leading order in this suppression factor, the right-hand side of Eq.~(\ref{pompevol}) only depends on the initial field $\zeta(0)$. For this reason, we do not use the full $\zeta(z)$ in the definition of $H(z)$, when it is used in the definitions of the semi-classical kernel functions.

\subsection{Perturbative approach}

Having concluded that approximations are required to solve the IPE in Eq.~(\ref{ipewig2}), we employ a perturbative approach, going beyond the semi-classical approximation. However, we use a different expansion parameter instead of the efficiency of the PDC process as the expansion parameter, because the expansion should be valid under conditions that allow highly efficient PDC.

If the magnitude of the parameter function of the pump beam (the power or average photon number of the pump state) is large enough, the region on which the Wigner functional of the pump beam is significantly different from zero is small compared to the distance from the origin of phase space. Hence, the contribution is insignificant unless $\beta$ is close to $\zeta$. So, one can use $\epsilon=\beta-\zeta$ as an expansion parameter for the perturbative approach. Although $\zeta$ varies is a function of $z$, the expansion parameter $\epsilon$ is bounded by the width of the Wigner functional of the pump beam, which does not vary, provided that the pump remains a pure coherent state. The semi-classical solution, which assumes $\beta=\zeta$, is the zeroth order term in the expansion. For higher orders, $\beta$ deviates from $\zeta$ by a small amount. In term of $F$, Eq.~(\ref{ipewig2}) becomes
\begin{widetext}
\begin{align}
i\hbar\partial_z F = & -i 2\hbar\int \left\{ \epsilon^*(\mathbf{k}) \left[\partial_z\zeta(\mathbf{k},z)\right]
+\left[\partial_z\zeta^*(\mathbf{k},z)\right] \epsilon(\mathbf{k}) \right\}\ \dbar k \nonumber \\
& +2\int \left\{ \left(\delta_1 F\right) \alpha^*(\mathbf{k}_2) \widetilde{T}^*(\mathbf{k}_1,\mathbf{k}_2,\mathbf{k}_3,z) \zeta(\mathbf{k}_3,z)
-\zeta^*(\mathbf{k}_3,z) \widetilde{T}(\mathbf{k}_1,\mathbf{k}_2,\mathbf{k}_3,z) \alpha(\mathbf{k}_1) \left(\delta_2^* F\right) \right. \nonumber \\
& +\left(\delta_1 F\right) \alpha^*(\mathbf{k}_2) \widetilde{T}^*(\mathbf{k}_1,\mathbf{k}_2,\mathbf{k}_3,z) \epsilon(\mathbf{k}_3)
-\epsilon^*(\mathbf{k}_3)\widetilde{T}(\mathbf{k}_1,\mathbf{k}_2,\mathbf{k}_3,z) \alpha(\mathbf{k}_1) \left(\delta_2^* F\right) \nonumber \\
& +\tfrac{1}{4} \left[ \left(\delta_1 \delta_2 F\right) + \left(\delta_1 F\right) \left(\delta_2 F\right) \right] \widetilde{T}^*(\mathbf{k}_1,\mathbf{k}_2,\mathbf{k}_3,z) \epsilon(\mathbf{k}_3)
-\tfrac{1}{4}\epsilon^*(\mathbf{k}_3) \widetilde{T}(\mathbf{k}_1,\mathbf{k}_2,\mathbf{k}_3,z)
\left[ \left(\delta_1^* \delta_2^* F\right) + \left(\delta_1^* F\right) \left(\delta_2^* F\right) \right]\nonumber \\
& \left. + \alpha^*(\mathbf{k}_1) \alpha^*(\mathbf{k}_2) \widetilde{T}^*(\mathbf{k}_1,\mathbf{k}_2,\mathbf{k}_3,z) \epsilon(\mathbf{k}_3)
-\epsilon^*(\mathbf{k}_3) \widetilde{T}(\mathbf{k}_1,\mathbf{k}_2,\mathbf{k}_3,z) \alpha(\mathbf{k}_1) \alpha(\mathbf{k}_2) \right\}\ \dbar k_1\ \dbar k_2\ \dbar k_3 .
\label{ipewigprt}
\end{align}
\end{widetext}
We now use the ansatz in Eq.~(\ref{ansatz}), but with
\begin{align}
\begin{split}
A(z) = & A_0(z) + A_1(z)\diamond\epsilon + A_1^{\dag}(z)\diamond\epsilon^* , \\
B(z) = & B_0(z) + B_1(z)\diamond\epsilon + B_2^*(z)\diamond\epsilon^* , \\
B^*(z) = & B_0^*(z) + B_1^*(z)\diamond\epsilon^* + B_2(z)\diamond\epsilon ,
\end{split}
\label{pertab}
\end{align}
where $A_0(z)$ and $B_0(z)$ are the semi-classical solutions and $A_1(z)$, $B_1(z)$ and $B_2(z)$ represent the first order perturbations to be solved. The different perturbation orders are distinguished by an expansion in the number of $\epsilon$-factors, after substituting Eq.~(\ref{ansatz}) and Eq.~(\ref{pertab}) into Eq.~(\ref{ipewigprt}). As expected, the zeroth order perturbation produces the same equations as in Eq.~(\ref{vgees1}) for the semi-classical case.

The first order perturbation can be further separated into eight differential equations, based on the different combinations of $\{\alpha,\alpha^*\}$. The two equations that are independent of $\{\alpha,\alpha^*\}$ give the same solution for the pump parameter function that is obtained in Eq.~(\ref{pompevol}). The remaining six equations are
\begin{align}
\begin{split}
\partial_z A_1(z) = & \tfrac{1}{2} H^*(z)\diamond B_1(z) + \tfrac{1}{2} B_2(z)\diamond H(z) \\
& + E_0(z)\diamond S(z)\diamond B_0(z) , \\
\partial_z B_1(z) = & \tfrac{1}{2} H(z)\diamond A_1(z) + \tfrac{1}{2} A_1^T(z)\diamond H(z) \\
& + B_0(z)\diamond S(z)\diamond B_0(z) , \\
\partial_z B_2(z) = & \tfrac{1}{2} H^*(z)\diamond A_1^T(z) + \tfrac{1}{2} A_1(z)\diamond H^*(z) \\
& + E_0(z)\diamond S(z)\diamond E_0^T(z) ,
\end{split}
\label{vgees3}
\end{align}
and their complex conjugates, where the $\diamond$-contractions are associated with the down-converted fields only, $E_0(z)\equiv A_0(z)-\mathbf{1}$ and $S(z) \equiv i2 \widetilde{T}^*(z)/\hbar$. The unknown functions all depend on an additional wave vector associated with the pump field, which will eventually be contracted on $\epsilon(\mathbf{k}_3)$ or $\epsilon^*(\mathbf{k}_3)$.

To solve these equations, we follow the same approach used for the semi-classical case. We integrate all the equations in  Eqs.~(\ref{vgees1}) and (\ref{vgees3}) with respect to $z$. Then we perform progressive back substitutions, with the initial conditions $A_1(0)=B_1(0)=B_2(0)=0$ in addition to those for $A_0(z)$ and $B_0(z)$. Thus we obtain expansions in terms of integrals of contracted $H$-kernels. The leading order terms for these kernels are given by
\begin{widetext}
\begin{align}
\begin{split}
A_1(z) = & \int_0^z \left[\int_0^{z_1}\int_0^{z_2}\mathcal{Z}\left\{H^*(z_2)\diamond H(z_3)\right\}\ \text{d}z_3\ \text{d}z_2\right] \diamond S(z_1)\diamond \left[\int_0^{z_1} H(z_4)\ \text{d}z_4\right]\ \text{d}z_1 , \\
B_1(z) = & \int_0^z \left[\int_0^{z_1} H(z_3)\ \text{d}z_2\right] \diamond S(z_1)\diamond
\left[\int_0^{z_1} H(z_3)\ \text{d}z_3\right]\ \text{d}z_1 , \\
B_2(z)  = & \int_0^z \left[\int_0^{z_1}\int_0^{z_2}\mathcal{Z}\left\{H^*(z_2)\diamond H(z_3)\right\}\ \text{d}z_3\ \text{d}z_2\right] \diamond S(z_1)\diamond \left[\int_0^{z_1}\int_0^{z_4}\mathcal{Z}\left\{H(z_4)\diamond H^*(z_5)\right\}\ \text{d}z_5\ \text{d}z_4\right]\ \text{d}z_1 .
\end{split}
\label{hoogkern}
\end{align}
\end{widetext}

Given the solutions for all these kernel functions, we obtain an expression for the state produced by the PDC process beyond the semi-classical approximation. We substitute Eq.~(\ref{pertab}) into the ansatz in Eq.~(\ref{ansatz}) and replace $\epsilon\rightarrow\beta-\zeta$. Then we combine it with the Wigner functional of the pump. The result is the second-order perturbative solution for the Wigner functional of the total state produce during the PDC process. It reads
\begin{align}
W_{\hat{\rho}} \approx &\ \mathcal{N}_0^2 \exp\left\{ -2\|\beta-\zeta(z)\|^2-\tau(z) \right. \nonumber \\
& \left. -\gamma^{\dag}(z)\diamond[\beta^*-\zeta^*(z)]-[\beta-\zeta(z)]\diamond\gamma(z)\right\} ,
\label{oplprt}
\end{align}
where the $\diamond$-contractions are with respect to the pump wave vector, and
\begin{align}
\begin{split}
\tau(z) \equiv &\ 2\alpha^*\diamond A_0(z)\diamond\alpha+\alpha\diamond B_0(z)\diamond\alpha \\
& +\alpha^*\diamond B_0^*(z)\diamond\alpha^* , \\
\gamma(z) \equiv &\ 2\alpha^*\diamond A_1(z)\diamond\alpha+\alpha\diamond B_1(z)\diamond\alpha \\
& +\alpha^*\diamond B_2(z)\diamond\alpha^* ,
\end{split}
\label{taugamma}
\end{align}
with the $\diamond$-contractions being with respect to the down-converted wave vectors.

The down-converted state observed in PDC experiments usually does not include the pump. Therefore, we trace out the pump degrees of freedom by performing the functional integration over $\beta$. In the process, all the terms that explicitly contain the parameter function $\zeta(z)$ are removed. It remains implicit in the definition of $H(z)$. The result after the integration is
\begin{equation}
W_{\hat{\sigma}} = \int W_{\hat{\rho}}\ \Dcirc[\beta] = \mathcal{N}_0 \exp\left(-\tau + \tfrac{1}{2}\gamma^{\dag}\diamond\gamma\right) ,
\label{troplprt}
\end{equation}
where the $\diamond$-contraction is with respect to the pump wave vector. The second term in the exponent is fourth order in $\{\alpha,\alpha^*\}$. Hence, the expression is not in Gaussian form.

\section{\label{applic}Applications}

The formalism presented here is intended for those applications of parametric down-conversion where both the particle-number degrees of freedom and the spatiotemporal degrees of freedom play important roles. Such applications are inevitably of a rather complex nature. While the presented formalism provides the means to incorporate all the relevant degrees of freedom, it does not completely remove the complexity of these calculations. As a result, an explicit example that demonstrates the power of this formalism is beyond the scope of the current paper. However, we present here a symbolic calculation that shows how this formalism would produce results that can go beyond the traditional semi-classical predictions.

Consider an observable $\hat{X}$ that represents a measurement applied to the down-converted state. If the Wigner functional of the observable is represented by $W_{\hat{X}}[\alpha]$, the result of the measurement would be given by
\begin{equation}
\langle \hat{X} \rangle = \tr\{\hat{\rho}\hat{X}\} = \int W_{\hat{X}}[\alpha] W_{\hat{\rho}}[\alpha]\ \Dcirc[\alpha] ,
\end{equation}
where $W_{\hat{\rho}}[\alpha]$ represents the Wigner functional of the down-converted state. For the sake of tractibility, we'll used the expression in Eq.~(\ref{oplprt}), prior to the functional integration over the pump field variable $\beta$.

Often such an observable would consist of {\em selective number operators}, representing detectors employed to observe single-photons or the photon number of the state at a specific location. The Wigner functional of such a selective number operator is given by a second-order polynomial functional:
\begin{equation}
W_{\hat{n}}[\alpha] = \alpha^*\diamond D\diamond\alpha-\tfrac{1}{2}\tr\{D\} ,
\end{equation}
where $D$ is a kernel function that models the experimental conditions imposed by the detector. It also incorporates the model for any optical system that directs the light toward the detector.

Such a second-order polynomial functional can be converted to a generating functional by adding it, multiplied with a generating parameter, to the exponent of the Wigner functional of the state. (If there are more than one detector, their Wigner fucntionals are included separately, each multiplied by its own generating parameter.) The resulting expression for the measurement becomes
\begin{equation}
\langle \hat{X} \rangle = \left. \partial_J \int \exp\left( F+J W_{\hat{X}} \right)\ \Dcirc[\alpha,\beta] \right|_{J=0} ,
\end{equation}
where $F$ is the polynomial functional exponent of the complete state. It is second order in $\{\alpha,\alpha^*\}$. Therefore, one can evaluate the functional integral over $\alpha$, leading to an expression of the form
\begin{align}
\langle \hat{X} \rangle = & \int \exp\left( -2\|\beta-\zeta\|^2 \right) \left. \partial_J P[\beta](J) \right|_{J=0}\ \Dcirc[\beta] \nonumber \\
 = & \int \exp\left( -2\|\beta-\zeta\|^2 \right) G[\beta]\ \Dcirc[\beta] ,
\label{verwx}
\end{align}
where $P[\beta](J)$ usually includes a functional determinant in the denominator and where we applied the derivative with respect to the generating parameter and set it to zero. In general, the functional integration over $\beta$ would not be tractable. However, we can proceed with the same perturbative approach used above to expand $G[\beta]$ as
\begin{align}
G[\beta] = &\ G_0 + (\beta^*-\zeta^*)\diamond G_1 + G_1^*\diamond (\beta-\zeta) \nonumber \\
& + (\beta^*-\zeta^*)\diamond G_{2a}\diamond (\beta-\zeta) \nonumber \\
& + (\beta-\zeta)\diamond G_{2b}\diamond (\beta-\zeta) \nonumber \\
& + (\beta^*-\zeta^*)\diamond G_{2b}\diamond (\beta^*-\zeta^*) + ... .
\label{gpert}
\end{align}
After substituting this expansion into Eq.~(\ref{verwx}), one can evaluate the functional integration over $\beta$ to obtain
\begin{equation}
\langle \hat{X} \rangle \approx G_0 + G_{2a} .
\end{equation}
The other terms in Eq.~(\ref{gpert}) fall away. The first term $G_0$ is the result that would be obtained from a semi-classical analysis. The second term $G_{2a}$ represents the contribution beyond the semi-classical approximation. It would be responsible for deviations in the experimental observations when compared to predictions that are purely based on a semi-classical analysis.

\section{\label{concl}Conclusions}

Using a Wigner functional approach, we obtained an evolution equation for the Wigner functional of the state produced during PDC in a second-order nonlinear medium under type I phase matching. For this purpose, we followed an infinitesimal propagation approach, in which the infinitesimal propagation operator is obtained from a comparison between the dynamical equations for the field operators and the commutators of these field operators with an ansatz for the infinitesimal propagation operator. The Wigner functional of the infinitesimal propagation operator then leads to the evolution equation for the Wigner functional of the state.

We then solved the evolution equation to obtain an expression for the Wigner functional of the state in the form of an exponential with a polynomial functional in its argument. It is done with the aid of a perturbative approach up to sub-leading order in the expansion parameter. The semi-classical solution is the zeroth order contribution in the polynomial functional. In this way, we obtained a solution beyond the semi-classical approximation. It allows one to consider the case where the pump suffers depletion during the PDC process. We also provided a generic discussion of applications where measurements are performed on the down-converted state to show how the contribution beyond the semi-classical approximation would affect the measurement results.

The fact that the final expression for the Wigner functional of the state obtained from the perturbative approach is not a Gaussian functional, has detrimental consequences for calculations using this result. However, all the kernel functions involved in the higher order terms in the exponent are suppressed, because they contain a vertex without the enhancement given by the parameter function. Therefore, one can use the more traditional perturbative approach, based on the low efficiency of the unenhanced PDC process to perform computations with this state. As we showeb generically, the expression for the down-converted state can be used with the aid of such perturbative methods to compute the measurement results expected in experiments involving such states. Therefore, we expect these results to be of significant relevance in applications, such as quantum metrology.

\section*{Acknowledgement}

This work was supported in part by funding from the National Research Foundation of South Africa (Grant Numbers: 118532).

\appendix

\section{\label{obv}Optical beam variables}

Fourier domain integrals for optical fields run over only three of the four quantities $\omega$, $k_x$, $k_y$ and $k_z$; by specifying any three of these quantities, the fourth is fixed by the dispersion relation $\omega = c|\mathbf{k}|$. Often, the three integration variables are chosen to be $k_x$, $k_y$ and $k_z$, which covers the three-dimensional space of propagating plane waves. It gives a symmetric three-dimensional representation of the field suitable for a Lorentz covariance formulation.

On the other hand, when dealing with optical beams in experiments, one may find it more convenient to integrated over $\omega$, $k_x$ and $k_y$ and thereby fix $k_z$ as
\begin{equation}
k_z = \sqrt{\frac{\omega^2}{c^2}-k_x^2-k_y^2} .
\label{kzinw}
\end{equation}
The set $\{\omega, k_x, k_y\}$ is referred to as the {\em optical beam variables}. It allows a natural setting in terms of which one can impose a monochromatic assumption. The spectrum of plane waves is restricted to a half-sphere, centered on the general direction of propagation of the optical beam, which is denoted as the $z$-axis. The plane waves that form part of the optical beam are those that have positive $k_z$-components. The transverse part of the wave vector is denoted by
\begin{equation}
\mathbf{K}=k_x \vec{x} + k_y \vec{y} .
\label{transk}
\end{equation}

Optical beam variables assumes a fixed propagation direction. As such, it breaks Lorentz covariance explicitly, but it is doing so in a way that is consistent with a Lorentz covariance formulation of the theory.

The definitions of all the quantities can be converted into equivalent definitions in terms of optical beam variables by changing the integration variables in the Fourier domain integrals from an integration over $k_z$ to an integration over $\omega$, using Eq.~(\ref{kzinw}). It produces the following transformation of the integration variable
\begin{equation}
\text{d}k_z = \frac{\omega\text{d}\omega}{c^2 k_z} ,
\label{dkznadw}
\end{equation}
which means that the Fourier domain measure is converted as follows
\begin{equation}
\frac{\text{d}^3 k}{(2\pi)^3 \omega} \rightarrow \frac{1}{c^2} \frac{\text{d}^2 k\ \text{d}\omega}{(2\pi)^3 k_z} \equiv \frac{1}{c^2} \dbar k.
\label{dk3nabv}
\end{equation}

The commutation relations of the ladder operators and the orthogonality conditions of the single-photon momentum basis are affected by their redefinitions in terms of optical beam variables. Using the properties of the Dirac-$\delta$ function, together with Eq.~(\ref{dkznadw}), we have
\begin{equation}
\delta(k_{z}-k_{z}') =  \delta(\omega-\omega') \frac{c^2 k_{z}}{\omega} .
\label{diracwkz}
\end{equation}
Additional factors of the speed of light in vacuum $c$ appear in Eqs.~(\ref{dk3nabv}) and (\ref{diracwkz}). These factors will appear in all the orthogonality conditions, commutation relations and Fourier domain integrals, unless we redefine the ladder operators and single-photon momentum basis in such a way that these factors of $c$ cancel. Therefore, we define a {\em momentum-frequency} basis in terms of the original momentum basis (ignoring the spin) as
\begin{equation}
\ket{\mathbf{k}} \equiv \ket{\mathbf{K},\omega} c
\label{twdef}
\end{equation}
removing a factor of $c$ from these states. The orthogonality condition for the momentum-frequency basis is
\begin{equation}
\braket{\mathbf{K},\omega}{\mathbf{K}',\omega'} = (2\pi)^3 k_{z} \delta(\mathbf{K}-\mathbf{K}') \delta(\omega-\omega') .
\label{inprodk2}
\end{equation}
The mapping from $\ket{\mathbf{k}}$ to $\ket{\mathbf{K},\omega}$ is one-to-one, provided that $k_z\geq 0$ and $\omega\geq 0$.

The momentum-frequency basis elements are generated (or destroyed) by associated ladder operators
\begin{equation}
\ket{\mathbf{K},\omega} = \hat{a}^{\dag}(\mathbf{K},\omega) \ket{\text{vac}} ,
\end{equation}
obeying the commutation relation
\begin{equation}
\left[\hat{a}(\mathbf{K},\omega),\hat{a}^{\dag}(\mathbf{K}',\omega')\right]
= (2\pi)^3 k_{z} \delta(\mathbf{K}-\mathbf{K}') \delta(\omega-\omega') .
\label{commutmf}
\end{equation}
Hence, the momentum-frequency ladder operators are related to the original ladder operators by
\begin{equation}
\hat{a}^{\dag}(\mathbf{k}) = c\ \hat{a}^{\dag}(\mathbf{K},\omega) .
\end{equation}

\section{\label{csap}Coherent state assisted approach}

It is often convenient to use identity operators resolved in terms of coherent states, as given by
\begin{equation}
\int \ket{\alpha} \bra{\alpha}\ \Dcirc[\alpha] = \mathds{1} ,
\label{volkoh}
\end{equation}
where
\begin{equation}
\Dcirc[\alpha] \equiv \mathcal{D}[q]\ \mathcal{D}\left[\frac{p}{2\pi}\right] ,
\label{alfamaat}
\end{equation}
to compute the Wigner functionals for a state or operator. For the case with an operator $\hat{A}$, we have
\begin{align}
W_{\hat{A}}[q,p] & = \int \braketa{q+\tfrac{1}{2}x}{\alpha_1} \bra{\alpha_1}\hat{A}\ket{\alpha_2}
\braketb{\alpha_2}{q-\tfrac{1}{2}x} \nonumber \\
& \times \exp(-i p\diamond x)\ \mathcal{D}[x]\ \Dcirc[\alpha_1,\alpha_2] ,
\label{wigalpha}
\end{align}
where $\alpha_1$ and $\alpha_2$ are the parameter functions of the fixed-spectrum coherent states.

Next, we use
\begin{align}
\braket{q}{\alpha} = & \pi^{-\Omega/4}\exp\left(-\tfrac{1}{2} q\diamond q -\tfrac{1}{2} \alpha_0^*\diamond\alpha_0 \right. \nonumber \\
& \left. + \sqrt{2} q\diamond\alpha_0-\tfrac{1}{2} \alpha_0\diamond\alpha_0\right) ,
\label{oorvqkoh}
\end{align}
to evaluate the overlaps in Eq.~(\ref{wigalpha}). The functional integration over $x$ then leads to
\begin{align}
W_{\hat{A}}[\alpha] = & \mathcal{N}_0 \int \exp\left(-2\|\alpha\|^2+2\alpha^*\diamond\alpha_1
+2\alpha_2^*\diamond\alpha\right. \nonumber \\
& \left. -\tfrac{1}{2}\|\alpha_1\|^2 -\tfrac{1}{2}\|\alpha_2\|^2 -\alpha_2^*\diamond\alpha_1 \right) \nonumber \\
& \times \bra{\alpha_1}\hat{A}\ket{\alpha_2}\ \Dcirc[\alpha_1,\alpha_2] ,
\label{asscohwig}
\end{align}
where we expressed the result in terms of $\alpha(\mathbf{k})$, instead of $q(\mathbf{k})$ and $p(\mathbf{k})$, and defined $\mathcal{N}_0\equiv 2^{\Omega}$. Evaluating the overlap $\bra{\alpha_1}\hat{A}\ket{\alpha_2}$ in Eq.~(\ref{asscohwig}) and performing the functional integrations over $\alpha_1$ and $\alpha_2$, one can obtain the Wigner functional for an arbitrary operator $\hat{A}$.


\end{document}